\documentclass[preprint,superscriptaddress,aps]{revtex4}

\usepackage{graphicx}
\usepackage{dcolumn}
\usepackage{bm}

\begin{document}

\preprint{}

\title{Nuclear charge radius of $^8$He}

\author{P.\ Mueller}
\email{pmueller@anl.gov} \affiliation{Physics Division, Argonne
National Laboratory, Argonne, Illinois 60439, USA}
\author{I.\ A.\ Sulai}
\affiliation{Physics Division, Argonne National Laboratory, Argonne, Illinois 60439, USA}
\affiliation{Department of Physics and Enrico Fermi Institute, University of Chicago, Chicago, Illinois 60637, USA}
\author{A.\ C.\ C.\ Villari}
\affiliation{GANIL (IN2P3/CNRS-DSM/CEA), B.P. 55027 F-14076 Caen Cedex 5, France}
\author{J.\ A.\ Alc\'antara-N\'u\~nez}
\affiliation{GANIL (IN2P3/CNRS-DSM/CEA), B.P. 55027 F-14076 Caen Cedex 5, France}
\author{R. Alves-Cond\'e}
\affiliation{GANIL (IN2P3/CNRS-DSM/CEA), B.P. 55027 F-14076 Caen Cedex 5, France}
\author{K.\ Bailey}
\affiliation{Physics Division, Argonne National Laboratory, Argonne, Illinois 60439, USA}
\author{G.\ W.\ F.\ Drake}
\affiliation{Physics Department, University of Windsor, Windsor, Ontario, Canada N9B 3P4}
\author{M. Dubois}
\affiliation{GANIL (IN2P3/CNRS-DSM/CEA), B.P. 55027 F-14076 Caen Cedex 5, France}
\author{C. El\'eon}
\affiliation{GANIL (IN2P3/CNRS-DSM/CEA), B.P. 55027 F-14076 Caen Cedex 5, France}
\author{G. Gaubert}
\affiliation{GANIL (IN2P3/CNRS-DSM/CEA), B.P. 55027 F-14076 Caen Cedex 5, France}
\author{R.\ J.\ Holt}
\affiliation{Physics Division, Argonne National Laboratory, Argonne, Illinois 60439, USA}
\author{R.\ V.\ F.\ Janssens}
\affiliation{Physics Division, Argonne National Laboratory, Argonne, Illinois 60439, USA}
\author{N. Lecesne}
\affiliation{GANIL (IN2P3/CNRS-DSM/CEA), B.P. 55027 F-14076 Caen Cedex 5, France}
\author{Z.-T.\ Lu}
\affiliation{Physics Division, Argonne National Laboratory, Argonne, Illinois 60439, USA}
\affiliation{Department of Physics and Enrico Fermi Institute, University of Chicago, Chicago, Illinois 60637, USA}
\author{T.\ P.\ O'Connor}
\affiliation{Physics Division, Argonne National Laboratory, Argonne, Illinois 60439, USA}
\author{M.-G.\ Saint-Laurent}
\affiliation{GANIL (IN2P3/CNRS-DSM/CEA), B.P. 55027 F-14076 Caen Cedex 5, France}
\author{J.-C.\ Thomas}
\affiliation{GANIL (IN2P3/CNRS-DSM/CEA), B.P. 55027 F-14076 Caen Cedex 5, France}
\author{L.-B.\ Wang}
\affiliation{Los Alamos National Laboratory, Los Alamos, New Mexico 87545, USA}

\date{\today}

\begin{abstract}
The root-mean-square (rms) nuclear charge radius of $^8$He, the
most neutron-rich of all particle-stable nuclei, 
has been determined for the first time to be 1.93(3)~fm. 
In addition, the rms charge radius of $^6$He was measured to be
2.068(11)~fm, in excellent agreement with a previous result.
The significant reduction in charge radius from
$^6$He to $^8$He is an indication of the 
change in the correlations of the excess neutrons and is
consistent with the $^8$He neutron halo structure.
The experiment was based on
laser spectroscopy of individual helium atoms cooled
and confined in a magneto-optical trap. Charge radii were extracted from the
measured isotope shifts with the help of precision atomic theory calculations.

\end{abstract}


\maketitle

Precision measurements of nuclear structure in light isotopes
are essential for a better understanding of nuclei and of
the underlying interactions between protons and neutrons.
{\it Ab~initio} calculations of light nuclei provide
quantitative predictions of nuclear properties based on empirical
nucleon-nucleon and three-nucleon interactions \cite{Pie01, Nav00}.
Investigations of very neutron-rich isotopes, among which the lightest are $^6$He ($t_{1/2} = 807$~ms)
and $^8$He ($t_{1/2} = 119$~ms), present especially stringent
tests for these calculations as they probe aspects of the interactions
that are less prevalent in nuclei closer to stability.

The differences in charge radii in the helium isotopes, where the two
protons are predominantly in a relative s-state, reflect primarily the
center-of-mass motion of the protons with respect to the neutrons. Therefore, the
charge radius is especially sensitive to neutron correlations.
The two excess neutrons in $^6$He and the four in $^8$He form a halo
with respect to a $^4$He core (or $\alpha$-particle) \cite{Tan92}.
The excess neutron pair in $^6$He is correlated in such a way that the recoil motion of the
core results in an increased charge radius \cite{Wan04}. In $^8$He, the four
excess neutrons are expected to be correlated in a more spherically symmetric way and
this recoil effect is expected to be smaller.

Here, we report on the first measurements of the nuclear charge radius of $^8$He.
Neutral helium atoms were laser cooled and confined in a magneto-optical trap, and
the isotope shift $\delta\nu_{A,A'}$ of an atomic transition between
isotopes $A$ and $A'$ was determined by laser spectroscopy. This isotope
shift can be expressed as:
\begin{equation}
\delta \nu _{A,A'} = \delta \nu^{\rm MS}_{A,A'} + K_{\rm FS} \; \delta \langle r^2 \rangle_{A,A'}.
\end{equation}
The mass shift $\delta \nu^{\rm MS}_{A,A'}$ and the field shift constant
$K_{\rm FS}$ of this two-electron system have both been precisely calculated
\cite{Dra04}, so that the change in mean-square nuclear charge radii $\delta
\langle r^2 \rangle_{A,A'}$ between the two 
isotopes can be extracted from the measured isotope shift. Combined with a
previous measurement on $^6$He \cite{Wan04} and
earlier studies of the two stable isotopes, $^3$He \cite{Shi95Mor06} and
$^4$He \cite{Sic82}, there is now a complete picture of the evolution of nuclear
charge radii in the helium isotopic chain.

The experiment was carried out at the GANIL
facility where $^6$He and $^8$He were simultaneously produced from a primary
beam of 75 MeV/u $^{13}$C impinging on a heated ($\sim 2000~$K) graphite target.
Low-energy (20 keV) beams of either $^6$He or 
$^8$He with yields of around $1\times 10^8$ and $5\times 10^5$ ions per
second, respectively \cite{Lan02}, were delivered to an adjacent low-radiation area where the
helium ion beam was stopped in a hot, 1~cm$^2$ sized graphite foil for neutralization.
The released neutral, thermal helium atoms were pumped within 250 ms into the atomic beam apparatus
resulting in rates of approximately $5\times 10^7$~s$^{-1}$ and $1 \times 10^5$~s$^{-1}$ for $^6$He and $^8$He, respectively.

The trapping and spectroscopy setup has been previously described in detail
in connection with the nuclear charge radius measurement
of $^6$He at Argonne's ATLAS facility \cite{Wan04,Wan04b}. The selective
cooling and trapping of helium atoms in a magneto-optical trap (MOT) was pivotal for
this work, providing single atom sensitivity, large signal-to noise ratios and
high spectroscopic resolution.
A beam of metastable helium atoms was produced through a LN$_2$-cooled gas
discharge. Transverse cooling and Zeeman slowing were applied to load the
metastable helium atoms of a selected isotope into the MOT.
Cooling and trapping were based on the cycling $2 ^3\!S_1 \rightarrow
2 ^3\!P_2$ transition at a wavelength of 1083~nm. The laser frequency for
this transition was controlled to an accuracy of 100~kHz to allow
reproducible switching between the respective isotopes. Detection and
spectroscopy of the atoms captured in the MOT were performed by exciting one
of the three cycling $2 ^3\!S_1 \rightarrow 3 ^3\!P_J$ transitions at 389~nm
and imaging the fluorescence light onto a photomultiplier tube. The frequency
of this probing laser was continuously measured relative to an iodine locked
reference laser. The capture and detection efficiency of the setup were
substantially improved compared to the previous $^6$He measurement
\cite{Wan04,Wan04b} by optimizing many components throughout the apparatus.
The signal-to-noise ratio of a single trapped atom reached $10$ within 50 ms of
integration time. The total capture efficiency was $1\times 10^{-7}$ and
yielded capture rates of around 20,000 $^6$He and 30 $^8$He atoms per hour.

\begin{figure}
\vspace{-0.7cm}
\includegraphics{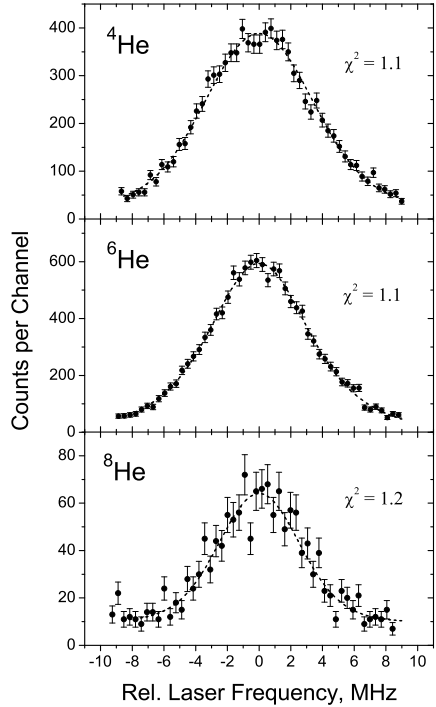}
\vspace{-1.3cm}
\caption{\label{fig:samplepeaks}Sample spectra for $^4$He, $^6$He and $^8$He taken
on the $2\,^3\!S_1 \rightarrow 3\,^3\!P_2$ transition at a probing laser intensity
of $\sim 3 \times I_{\rm sat}$. Error bars are statistical uncertainties, the dashed lines
represent least squares fits (with the listed reduced $\chi ^2$) using Voigt profiles.
The apparent peak broadening towards lower masses is due to the $m^{-1/2}$ scaling of
the residual Doppler width.}
\end{figure}

In the {\it capture} mode, the probing laser at 389~nm was tuned on
resonance for maximum fluorescence and the trapping laser detuning
and intensity were selected for highest capture efficiency. Single atom
detection of $^8$He triggered the system to switch into
the {\it spectroscopy} mode for 200~ms, while for $^4$He and $^6$He the system
was continuously switched to the {\it spectroscopy} mode with a rate of 2.5~Hz.
During the {\it spectroscopy} mode, the fluorescence rate was recorded as a function
of probing laser frequency by scanning the probing laser with a 83~kHz repetition
rate over $\pm$9~MHz relative to the respective resonance center. 
Additionally, the detuning and intensity of the trapping laser were
reduced, the slowing light was
turned off, and the probing and trapping laser beams were
alternately switched on and off with a 100~kHz repetition rate and a cycle
of 2~$\mu$s probing vs. 8~$\mu$s trapping. This scheme was developed to
eliminate AC Stark shifts caused by the trapping light while minimizing
systematic heating and cooling of the atoms caused by the probing light.

A total of twelve sets of measurements for the $^6$He-$^4$He isotope shift 
and eight for $^8$He-$^4$He
were taken during three days. The measurements for $^6$He and the reference
isotope $^4$He were each performed at several settings for the probing laser intensity:
from $\sim 3\times I_{\rm sat}$ down to $\sim 0.3\times I_{\rm sat}$,
where $I_{\rm sat} = 3.4$~mW/cm$^2$ is the saturation intensity of the
transition. Small systematic shifts of the resonance frequencies observed 
at high intensities showed no significant difference between $^4$He and $^6$He, but
were found to vary slightly over time, consistent with variations of the probing
laser beam alignment. This effect was taken into account for the $^8$He data, which
could only be recorded with signal-to-noise ratios sufficient
for single-atom detection, i.e., at $\sim 3\times I_{\rm sat}$.

Samples of resonance profiles for each isotope taken at $\sim 3 \times I_{\rm sat}$
are given in Fig.~\ref{fig:samplepeaks}. The $^8$He peak represents data
integrated over 60 individually trapped atoms accumulated in two hours, while the
$^4$He and $^6$He peaks were typically acquired in less than one minute. All
peaks could be fit well with Voigt profiles.
The Gaussian widths of the fitted Voigt profiles scale
with $m^{-1/2}$, as expected from a mass-independent temperature of the
trapped atoms, and become smaller at lower intensity of the probing laser.

\begin{figure}
\includegraphics{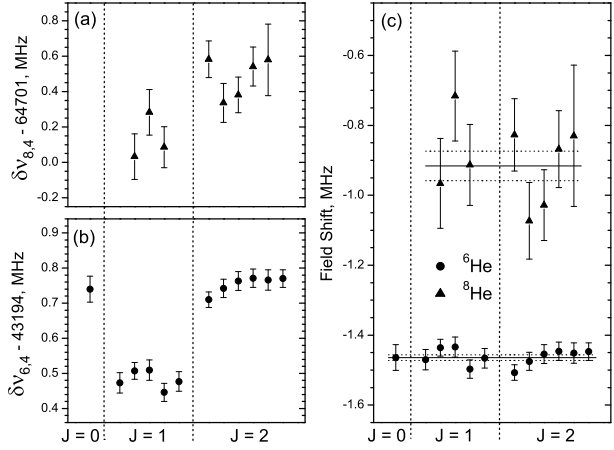}
\vspace{-0.7cm}
\caption{\label{fig:ISandFS}Experimental isotope shifts relative to $^4$He
  from the individual measurements for $^8$He (a) and
  $^6$He (b). As expected, the isotope shift depends on the $J$ of the upper
  $3\,^3\!P_J$ state. However, the extracted field shift values plotted
  in (c) show no systematic $J$ dependence for either isotope.
  The horizontal lines in (c) mark the weighted averages and statistical
  error bands of the field shift.}
\end{figure}

\begin{table}
\caption{Weighted averages of the experimental isotope shifts
  $\delta \nu _{A,4}$ (including recoil correction) for the different
  transitions in $^6$He and $^8$He. The field shift
  $\delta \nu^{\rm FS}_{A,4}= K_{\rm FS} \; \delta \langle r^2 \rangle_{A,4}$
  was calculated for each transition
  using the listed {\it theoretical} mass shift values $\delta \nu^{\rm MS}_{A,4}$.
  All values are in MHz. The errors given in 
  parentheses for $\delta \nu_{A,4}$ and $\delta \nu^{\rm FS}_{A,4}$ 
  include only statistical uncertainties.}
\label{tab:ISvalues}
\begin{tabular}{llr@{}lr@{}lr@{}l}
\hline
\multicolumn{2}{c}{Transition}&
\multicolumn{2}{c}{$\delta \nu _{A,4}$} &
\multicolumn{2}{c}{$\delta \nu^{\rm MS}_{A,4}$} &
\multicolumn{2}{c}{$\delta \nu^{\rm FS}_{A,4}$}\\
\hline
$^6$He & $2\,^3\!S_1 \rightarrow 3\,^3\!P_0$ & 43194&.740(37) & 43196&.204 & -1&.464(37) \\
       & $2\,^3\!S_1 \rightarrow 3\,^3\!P_1$ & 43194&.483(12) & 43195&.943 & -1&.460(12) \\
       & $2\,^3\!S_1 \rightarrow 3\,^3\!P_2$ & 43194&.751(10) & 43196&.217 & -1&.466(10) \\
\hline
$^8$He & $2\,^3\!S_1 \rightarrow 3\,^3\!P_1$ & 64701&.129(73) & 64701&.999 & -0&.870(73) \\
       & $2\,^3\!S_1 \rightarrow 3\,^3\!P_2$ & 64701&.466(52) & 64702&.409 & -0&.943(52) \\
\hline
\end{tabular}
\end{table}

The isotope shifts for $^6$He and $^8$He relative to $^4$He obtained in the
individual measurements are plotted in Fig.~\ref{fig:ISandFS} along with the
extracted field shifts. Table~\ref{tab:ISvalues} lists the weighted averages
of isotope shifts and field shifts separately for the different fine
structure levels $^3\!P_J$. The isotope shift for the $2\,^3\!S_1
\rightarrow 3\,^3\!P_2$ transition in $^6$He agrees with
the previously published value of 43194.772(33)~MHz \cite{Wan04} within the
quoted {\it statistical} uncertainties. The isotope shift values for the
different transitions in $^6$He show variations by 250~kHz, as predicted by
the atomic theory calculations.
The extracted field shifts for all three transitions agree well within
statistical uncertainties. This is a valuable consistency test for atomic
theory as well as a check for a class of systematic errors in the
experiment, since the strengths of these three transitions vary by a factor
of up to five. Hence, the field shifts over all three transitions in $^6$He
were averaged as independent measurements, and likewise for the two transitions
observed in $^8$He.

The final field shift results for both isotopes are listed in Table~\ref{tab:errors}
along with the contributions from statistical and systematic uncertainties. Besides
photon counting statistics, there are two additional random effects: the frequency drift of
the reference laser and variations in the power-dependent frequency shift
due to small drifts in the probing laser alignment. Both lead to significant
scattering of the results during the roughly two hour integration time
needed for each $^8$He measurement, but are insignificant
in the case of $^6$He.
A significant systematic uncertainty is caused by Zeeman shifts that might have varied 
among isotopes if the atoms were not located exactly at the zero B-field
position of the MOT. Limits on this effect are set conservatively at
$\le$30~kHz for the $^6$He-$^4$He isotope shift, and $\le$45~kHz for
$^8$He-$^4$He. Moreover, two corrections are applied to the measured
isotope shifts as listed in Table~\ref{tab:errors}: photon recoil and
nuclear polarization. The first was trivially and accurately calculated.
The latter depends on the nuclear polarizability, which
was extracted from measurements of the electric dipole 
strength \cite{Pac07,Iwa00}.
The uncertainty in the nuclear mass enters as an additional systematic
effect via the theoretical mass shift.
This effect is the single biggest contribution to the final uncertainty 
for $^8$He, but plays only a minor role for $^6$He.
Improved mass measurements for both isotopes are in preparation,
using Penning trap mass spectrometry \cite{Bla07Dil07}.

\begin{table}
\caption{Statistical and systematic uncertainties and corrections on the
  combined results for the field shifts of $^6$He and $^8$He relative to
  $^4$He. All values are in MHz.}
\label{tab:errors}
\begin{tabular}{l||r@{}l|r@{}l||r@{}l|r@{}l }
\hline
\multicolumn{1}{c||}{} &
\multicolumn{4}{c||}{$^6$He} &  
\multicolumn{4}{c}{$^8$He}\\
\multicolumn{1}{c||}{} &
\multicolumn{2}{c|}{value} &  
\multicolumn{2}{c||}{error} &  
\multicolumn{2}{c|}{value} &  
\multicolumn{2}{c}{error}\\
\hline
{\it Statistical} & & & & & & & & \\
Photon counting & & & 0&.008 & & & 0&.032 \\
Probing laser alignment & & & 0&.002  & & & 0&.012 \\
Reference laser drift & & & 0&.002  & & & 0&.024 \\
\hline
{\it Systematic}  & & & & & & & & \\
Probing power shift & & & & & & & 0&.015 \\
Zeeman shift & & & 0&.030  & & & 0&.045 \\
Nuclear mass & & & 0&.015 & & & 0&.074 \\
\hline
{\it Corrections}  & & & & & & & & \\
Recoil effect    & 0&.110 & 0&.000 & 0&.165 & 0&.000 \\
Nuclear polarization & -0&.014 & 0&.003 & -0&.002 &  0&.001 \\
\hline
\hline
$\delta \nu _{A,4}^{\rm FS}$ combined & -1&.478 & 0&.035 & -0&.918 & 0&.097\\
\hline
\end{tabular}
\end{table}

\begin{table}
\caption{Relative and absolute charge radii for all particle-stable helium isotopes. 
  The absolute $^3$He radius is calculated with the relative value from 
  Ref. \cite{Shi95Mor06} and the absolute $^4$He value from Ref. \cite{Sic82}.
  Values for $^6$He and $^8$He are from this work.}
\label{tab:chargeradii}
\begin{tabular}{l|c|c|c|c}
\hline
 & $^3$He & $^4$He & $^6$He& $^8$He\\
\hline
$\delta \langle r^2 \rangle_{A,4}$, fm$^2$ & 1.059(3) & -  & 1.466(34)& 0.911(95) \\ 
$\langle r^2 \rangle _{\rm ch}^{1/2}$, fm  & 1.967(7) & 1.676(8) & 2.068(11) & 1.929(26) \\
\hline
\end{tabular}\\
\end{table}

Table~\ref{tab:chargeradii} lists the final results for the difference in
mean-square charge radius of $^6$He and $^8$He relative to $^4$He, which
follow directly from the field shift using $K_{\rm FS}=1.008$~fm$^2/$MHz from
atomic theory \cite{Dra04}. The absolute charge radii for both isotopes are based on a
value of 1.676(8)~fm for the $^4$He charge radius \cite{Sic82}.
For a comparison of our results on rms charge radii $\langle r^2 \rangle
 _{\rm ch}^{1/2}$ to the rms {\it point-proton} radii $\langle r^2 \rangle _{\rm pp}^{1/2}$,
 typically quoted by theoretical papers, the relation $\langle r^2 \rangle_{\rm pp} =
 \langle r^2 \rangle _{\rm ch} - \langle R_{\rm p}^2 \rangle - \frac{3}{4 M_{\rm p}^2} -
 \frac{N}{Z}\langle R_{\rm n}^2 \rangle$
was used, which takes into account contributions from the mean-square charge radii of the
proton and neutron (with $\langle R_{\rm p}^2 \rangle = 0.769(12)$~fm$^2$ and $\langle R_{\rm n}^2
\rangle = -0.1161(22)$~fm$^2$ \cite{Yao06}) and the Darwin-Foldy term $\frac{3}{4M_{\rm p}^2} = 0.033$~fm$^2$ \cite{Fri97}. 
The effects of nuclear spin-orbit interaction and meson exchange currents,
expected to be on the order of or below the experimental uncertainties, are not taken into account
and will require further theoretical investigation.

\begin{figure}
\includegraphics{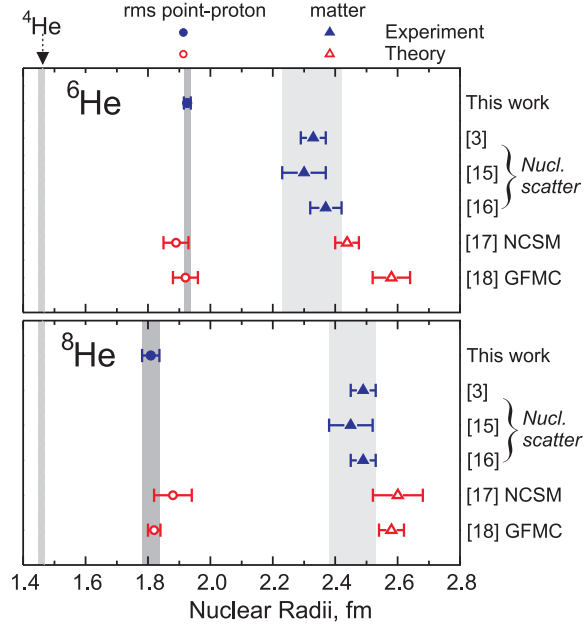}
\caption{\label{fig:compare}Comparison of rms point-proton radii (circles) and
  matter radii (triangles)
  for $^6$He and $^8$He between experiment (solid symbols) and
  theory (open symbols). The vertical bands represent {\em experimental} error
  bands, consistent with the spread and error bars of the reported values,
  and the $^4$He rms point-proton radius
  from \cite{Sic82}. For comparison, the $^3$He point-proton radius is 1.77(1)~fm \cite{Shi95Mor06}.}
\end{figure}

The experimental rms point-proton radii from this work are plotted in Fig.~\ref{fig:compare}
along with matter radii (i.e., point-nucleon radii) extracted from strong interaction cross section measurements
 \cite{Tan92,Alk97,Kis05}. While the latter are
model dependent, different methods give consistent matter radii. The matter radius for $^4$He should be
the same as the indicated point-proton radius. Also given in Fig.~\ref{fig:compare} are the values
from {\it ab~initio} calculations based on the no-core shell model (NCSM) \cite{Cau06} and
Green's function Monte Carlo (GFMC) techniques \cite{Pie07}. Apart from those, there are a number of cluster model
calculations providing values for rms point-proton and matter radii of both isotopes \cite{cluster}.

Most strikingly, the rms charge radius {\it decreases} significantly from $^6$He to $^8$He,
while the matter radius seems to {\it increase}.
The larger matter radius of $^8$He is consistent with there being more
neutrons and more nucleons altogether. On the other hand, the larger charge radius of ${^6}$He
is consistent with the interpretation that the two neutrons are correlated so that
on average they spend more time together on one side of the core rather than on opposite
sides. As a result, the recoil motion of the $\alpha$-like core against the correlated pair of
neutrons smears out the charge distribution. In ${^8}$He, the four excess
neutrons are distributed in a more spherically symmetric fashion in the halo and the
smearing of the charge in the core is correspondingly less, leading to a
{\em reduction} in the charge radius. These effects are reproduced rather well
by {\it ab initio} calculations, giving further confidence in our understanding of
nuclear forces and in the method of calculation.

\begin{acknowledgments}
We acknowledge the contribution of the GANIL accelerator (Service des Acc\'el\'erateurs) and physical
technical (Service Techniques de la Physique) staff. We thank J.\ P.\ Schiffer, S.\ Pieper and R.\ Wiringa
for stimulating discussions, and J.\ P.\ Greene, D.\ Henderson, S.\ M.\ Hu, C.\ L.\ Jiang, M.\ Notani, R.\ C.\ Pardo,
K.\ E.\ Rehm, B.\ Shumard and X.\ Tang for contributions in the early phase of this experiment.
This work was supported by the U.S. Department of Energy, Office of Nuclear Physics, under
Contract No. DE-AC02-06CH11357. G. Drake acknowledges support by NSERC and by SHARCnet.
Experiment performed at GANIL (IN2P3/CNRS - DSM/CEA).
\end{acknowledgments}

\end{document}